\documentclass[twocolumn,showpacs,preprintnumbers,amsmath,amssymb,floatfix]{revtex4}

\usepackage{epsfig}
\usepackage[rflt]{floatflt}
\newcommand{\kbf}{\mathbf{k}}
\newcommand{\ptot}{\pi_{\mathrm{tot}}}
\newcommand{\xbf}{\mathbf{x}}

\begin{document}

\title{Resolution Exchange Simulation}
\author{Edward Lyman\footnote{elyman@ccbb.pitt.edu}, 
F. Marty Ytreberg, and Daniel M. Zuckerman\footnote{dmz@ccbb.pitt.edu}}
\affiliation{Department of Computational Biology, School of Medicine  
and Department of Environmental and Occupational Health, 
Graduate School of Public Health, BST W1041, 200 Lothrop St., University
of Pittsburgh, Pittsburgh, PA 15261}
\date{\today}
\begin{abstract}
We extend replica exchange simulation in two ways, and apply our approaches to biomolecules.
The first generalization permits exchange simulation between models of differing resolution --- 
i.e., between detailed and coarse-grained models.
Such ``resolution exchange'' can be applied to molecular systems or spin systems.
The second extension is to ``pseudo-exchange'' simulations, which
require little CPU usage for most levels of the exchange ladder and 
also substantially reduces the need for overlap between levels. Pseudo exchanges can 
be used in either replica or resolution exchange simulations.
We perform efficient, converged simulations of a 50-atom peptide to illustrate the new approaches.\\
\\
\vspace{0.5cm}
Accepted for publication in physical review letters.
%
\end{abstract}

\maketitle

The simulation of biomolecules with $10^{4}-10^{5}$ degrees 
of freedom has become routine,  
thanks to the accessibility of powerful computing resources, the 
development of reliable simulation software, and standardized 
empirical potential energy 
functions. For many biological applications, such as binding free energy estimation, 
it is desirable to 
generate an equilibrated ensemble of conformations. 
In principle, standard Monte Carlo (MC) and molecular dynamics (MD) algorithms 
are perfectly ergodic, and therefore will eventually generate such ensembles. 
In practice, the $\mu\sec-\sec$ timescale, which describes biologically 
relevant fluctuations, is not within reach of computation even for small proteins.

Two broad strategies have been developed to address this problem. 
In one approach, 
dating to the 
earliest computational studies of proteins\cite{warshel-nature,go-folding75}, 
coarse-grained protein representations are adopted. This 
strategy continues to be popular\cite{ital-villin-fold1,zuckerman-calmod}. 

A second class of strategies attempts directly to enhance sampling 
of atomic-resolution models, including multiple time 
step methods\cite{berne-mts-md94,berne-mts-mc02}, replica exchange\cite{swendsen-repex}/parallel 
tempering\cite{nemoto-repex,hansmann-pt97,okamoto-repex-md}, and other generalized ensemble 
techniques\cite{depablo-dos-prl}. 
Parallel tempering (PT), which employs a ladder of replicas simulated at increasing temperatures, 
is widely used 
for state-of-the-art molecular dynamics simulations, but presently 
is limited to small proteins\cite{garcia-prl-big-repex-md}, as the resources required increase
rapidly with the system size. 
\begin{figure}[tp]
\epsfig{file=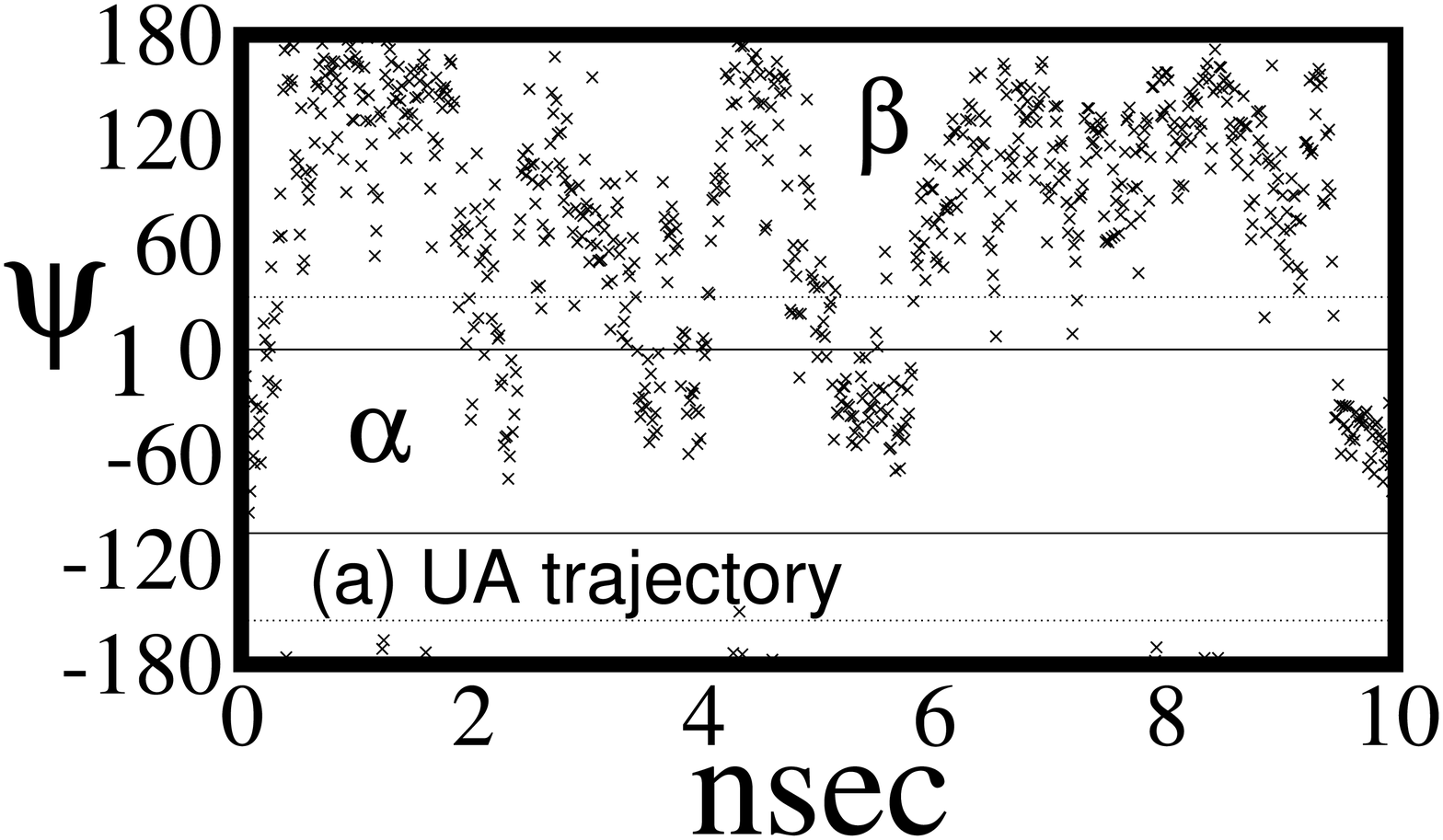, width=.46\columnwidth}
\epsfig{file=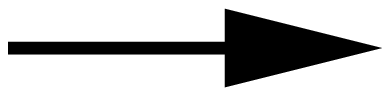, width=.04\columnwidth}
\epsfig{file=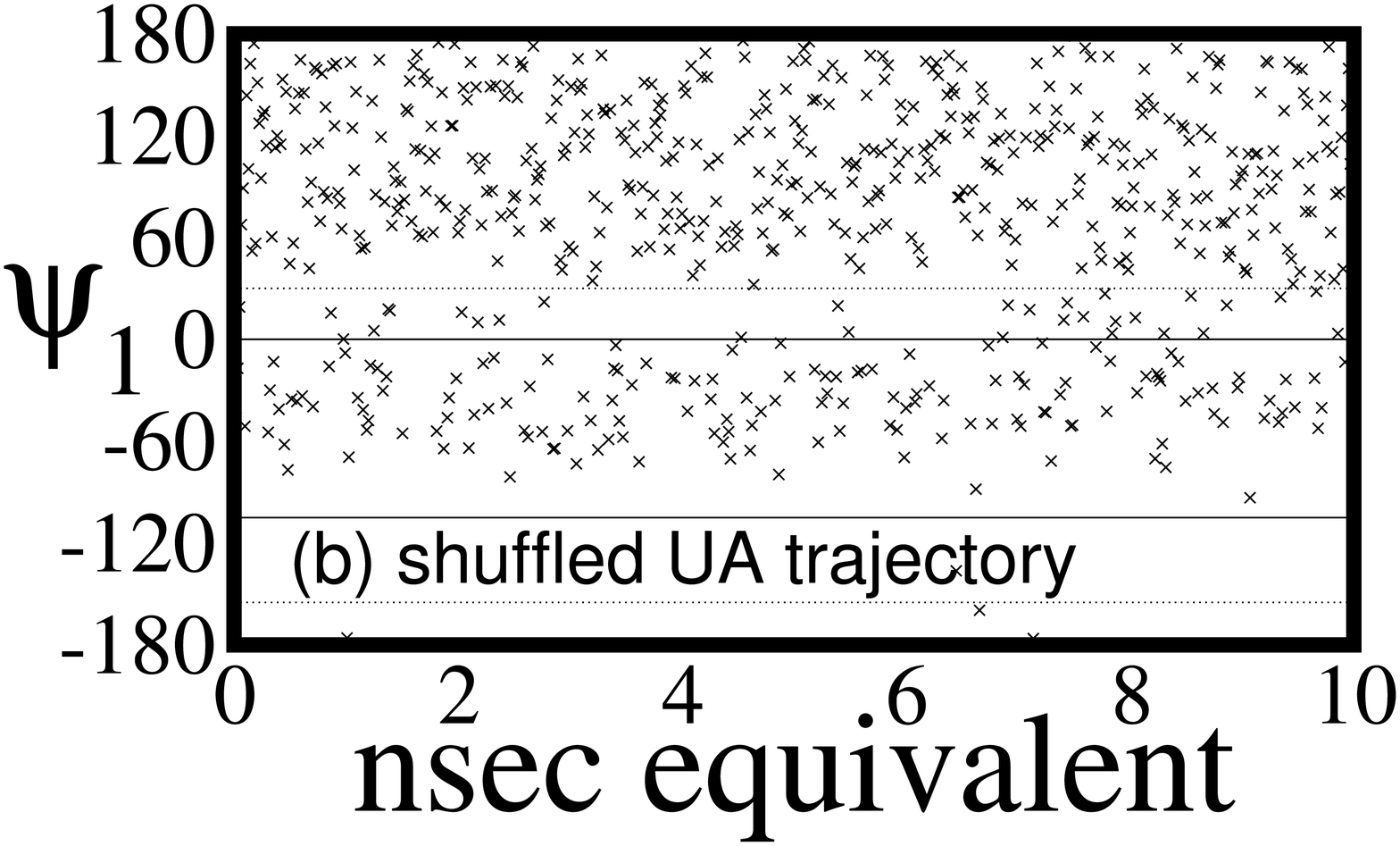, width=.46\columnwidth}
\hspace*{.6\columnwidth}
\epsfig{file=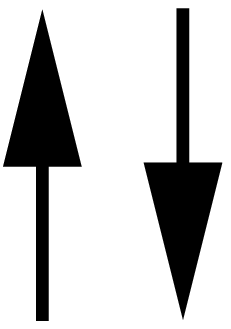, height=0.25cm}\\
\epsfig{file=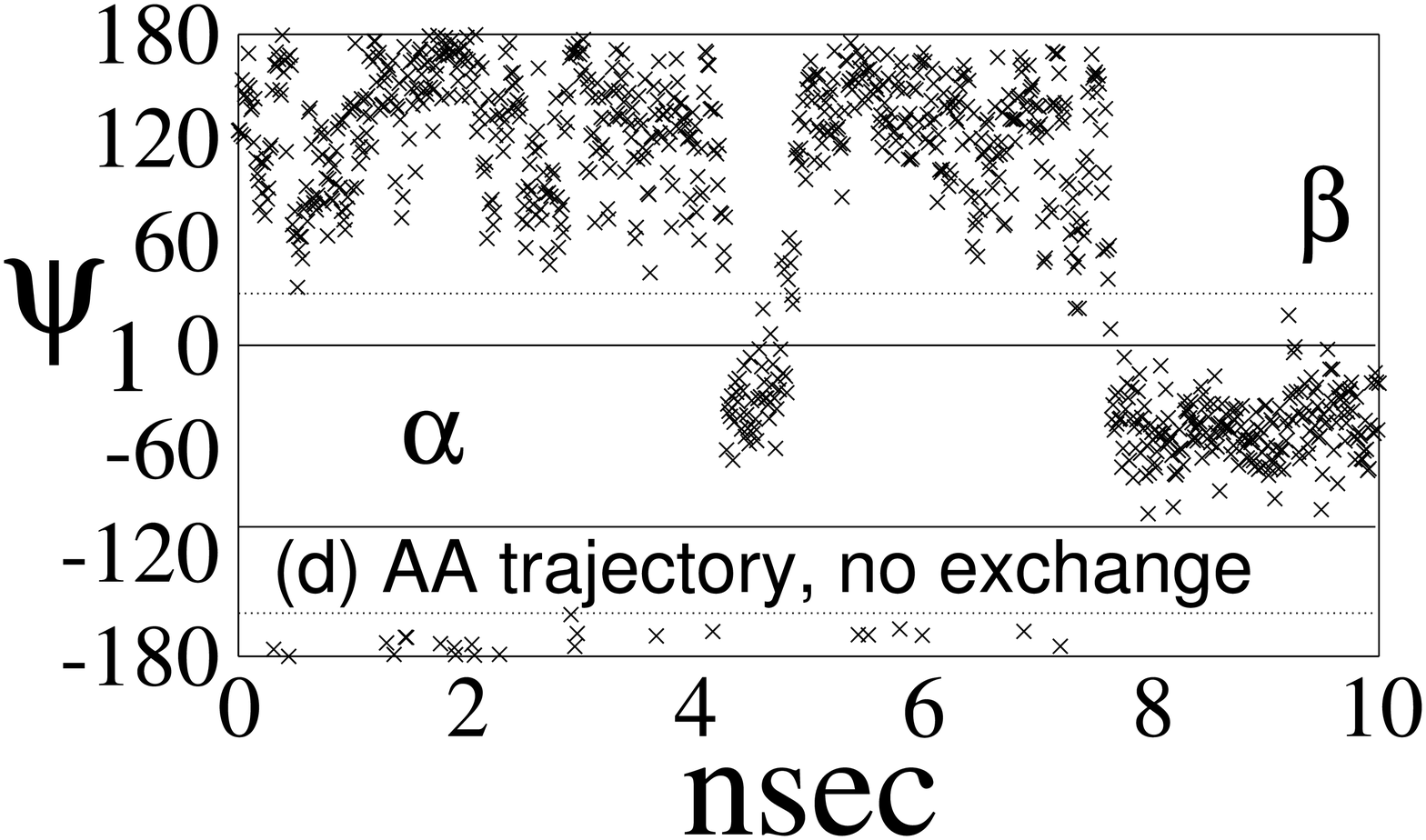, width=.46\columnwidth}
\hspace{.04\columnwidth}
\epsfig{file=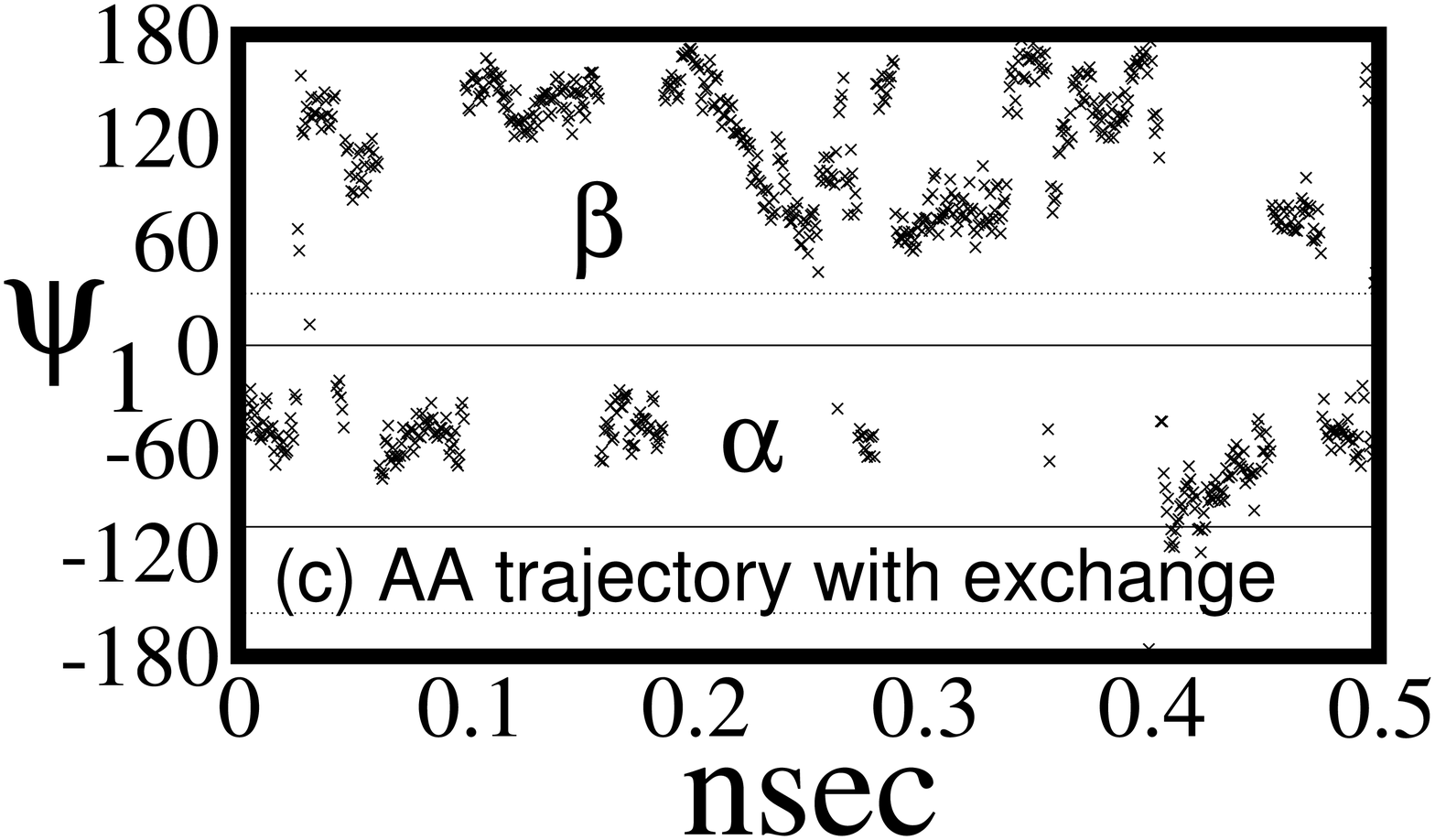, width=.46\columnwidth}
\caption{Coarse-grained simulation accelerates transitions 
in high-resolution simulation with res-ex and ps-ex. 
(a) United atom trajectory for dileucine peptide, 
showing transitions between $\alpha$ and $\beta$ states, which is randomly 
re-ordered to create trajectory (b) with extremely rapid transitions. Performing 
pseudo-exchanges with the shuffled trajectory, the res-ex algorithm generates the 
all-atom trajectory in (c). The time axes in (c) and (d) differ by a factor of $20$, 
underscoring the effectiveness of the ps-ex/res-ex protocol.
\label{fig1}
\vspace{-.7cm}
}
\end{figure}

This Letter presents two new tools for biomolecular simulation, by extending the PT approach 
and exploiting the speed of coarse-grained models.
The first extension is a ``resolution exchange'' (res-ex) algorithm which --- instead of 
using high-temperature simulation to increase sampling, as does PT --- uses inexpensive coarse-grained models to cross barriers.
Boltzmann-weighted ensembles are produced. 
The algorithm is implemented in close analogy to PT, and can also be applied to 
magnetic systems (e.g., the Ising model). 
The res-ex approach is natural for proteins, and indeed the kernel of the idea was suggested in the
early days of protein simulation\cite{warshel-nature}. More recently, the 
approach has been implemented in an \emph{ad hoc} way, without proper statistical 
weighting \cite{ital-villin-fold1}. A rigorous method to calculate free energy differences  
between all-atom and coarse-grained models was demonstrated by Warshel and 
coworkers\cite{warshel-coarse-fold99}. 

Our res-ex approach is conceptually related to work on Ising systems by Brandt and 
coworkers (e.g., \cite{brandt-ron-rmg,brandt-etal-prl-invmcrg}). The res-ex approach 
is distinguished, however, by its simplicity, its ready applicability to biomolecules, 
and the ability to employ arbitrary coarse-grained Hamiltonians --- rather than truly 
renormalized Hamiltonians, as in \cite{brandt-ron-rmg,brandt-etal-prl-invmcrg}. 

We also introduce ``pseudo-exchange'' (ps-ex) processes which should 
significantly improve the efficiency of  
\emph{any type of exchange simulation}, whether one swaps temperatures (as in PT), 
Hamiltonians \cite{okamoto-multidim-repex} or resolution (res-ex).  
Pseudo-exchanges are performed between a simulation in progress and one 
which has already been completed.
The key advantage of ps-ex is that it permits uneven distribution of CPU time 
among levels of the exchange ladder.
Because all exchange simulations are limited by the sampling obtained at the 
highest level --- i.e., highest temperature (for PT) or lowest resolution (res-ex) --- the 
bulk of CPU time should be devoted to this top level. 
Although an uneven distribution of CPU time among levels (replicas) would be 
awkward in a truly parallel implementation, it is natural and highly efficient in a serial ps-ex simulation.
Furthermore, there is essentially no disadvantage to multiple 
independent runs, as compared to a single parallel simulation.



\emph{Resolution Exchange Theory.} 
The key idea behind res-ex is that, in addition to swapping temperature 
labels (PT) or parameters of the Hamiltonian \cite{okamoto-multidim-repex}, one can also swap 
a \emph{subset} of configurational coordinates.
A well-chosen subset of coordinates of a detailed model can comprise the 
full set of coordinates for a coarse-grained model, as we demonstrate below.

A general exchange process is constructed by considering two \emph{independent} 
simulations of a protein (or a spin system) carried out in parallel, each sampling its own 
distribution $\pi_1$ or $\pi_2$. 
In common cases, the distributions will be given by
$\pi_i(\Phi_i,\xbf_i;\mathbf{k}_i;T_i) = \exp{(-U(\Phi_i,\xbf_i;\kbf_i) / k_B T_i)}$, 
where a configuration is composed of coordinates $\{\Phi,\xbf\}$ which include 
an arbitrarily chosen ``coarse'' subset $\Phi$, and
where $\mathbf{k}$ denotes the parameters of the potential function $U$ and $k_B T$ is the product of Boltzmann's constant and the temperature. 
A general exchange process consists of a swap of a set of the arguments of the $\pi$ functions:
swapping $T_1 \leftrightarrow T_2$ leads to PT, and swapping $\kbf_1 \leftrightarrow \kbf_2$ values leads to Hamiltonian exchange.
To achieve resolution exchange, one can swap values of the set of 
coarse coordinates, $\Phi_1 \leftrightarrow \Phi_2$, noting that the 
corresponding potential parameters $\kbf_\Phi$ need not match in the two systems. 
It is indeed possible to swap an arbitrary combination of coordinates and parameters.

Specializing, for clarity, to resolution exchange, we consider 
independent simulations governed by a ``high resolution'' potential function
$U_{H}(\{\Phi,\mathbf{x}\})$ and a coarse-grained (low-resolution) potential $U_{L}(\{\Phi\})$.
Occasionally,
we attempt an exchange move by swapping the $\{\Phi\}$ subset. 
The set $\{\Phi,\mathbf{x}\}$ may be, for example, all the atomic coordinates of a protein, 
while the subset $\{\Phi\}$ may be only the coordinates of the backbone. 
For a spin system, $\{\Phi\}$ may correspond to a block spin, and $\mathbf{x}$ 
to the orientations of the local spins relative to the block spin.  
 
To develop the exchange criterion, assume that at an exchange point the system 
is characterized by a high-resolution configuration $\{\Phi_{a}, \mathbf{x}_{a}\}$
and a low-resolution configuration $\{\Phi_{b}\}$. Attempting to exchange the
$\{\Phi\}$ subset yields the trial conformations $\{\Phi_{b}, \mathbf{x}_{a}\}$
and $\{\Phi_{a}\}$. 
Because the simulations are independent, the weight of the composite system is 
given by the simple product $\ptot = \pi_1 \pi_2 = \pi_H \pi_L$, and
detailed balance will be satisfied if we accept such moves with a Metropolis rate $\min[1,R]$, 
where $R$ is given by 
\begin{equation}
R=\frac
{\pi_{1}(\text{new})\pi_{2}(\text{new})}
{\pi_{1}(\text{old})\pi_{2}(\text{old})}
=\frac
{\pi_{H}(\Phi_{b},x_{a})\pi_{L}(\Phi_{a})}
{\pi_{H}(\Phi_{a},x_{a})\pi_{L}(\Phi_{b})}
. 
\label{para_rate}
\end{equation}
The analogy to PT and Hamiltonian exchange is clear, but we have now extended the approach.

Naturally, there are limitations on the types of models which can be successfully exchanged, much as PT temperature increments are limited.
In the results presented below, we successfully performed exchanges between all-atom and united-atom models of a peptide.

\emph{Pseudo-exchange simulation.} 
The res-ex and PT algorithms are motivated by the likelihood 
that the ``top level'' simulation (i.e., lowest resolution or highest $T$) will more rapidly 
cross barriers and converge to an equilibrium ensemble of conformations. 
While the associated convergence time is expected to be quite long, even for the top level, 
it is far from clear that the attainment of an equilibrium ensemble at a lower level 
requires the same length of simulation.
Indeed, given that barriers should be crossed many times at the top level, 
\emph{signficantly less simulation time should be required at the lower levels of the exchange ladder.}
Our results show this to be true.
Yet it would seem impossible to allot \emph{a priori} appropriate CPU resources among the 
various ladder levels in a conventional parallel exchange simulation.

A ``pseudo exchange'' process is the key to efficiently distributing computing time among ladder levels.
The first step is to generate a well-sampled ensemble at the top level (highest 
temperature or lowest resolution) and randomly re-order this trajectory(FIG.\ref{fig1}a-\ref{fig1}b).
While such shuffling preserves the distribution of states of the original trajectory, the 
shuffled trajectory exhibits a feature key for exchange simulation: \emph{extremely rapid 
barrier hops,} as in FIG.\ref{fig1}b.

One now performs a ps-ex simulation with the shuffled trajectory.
As with conventional exchange, one runs an independent lower level simulation (FIG.\ref{fig1}c), but 
now exchanges are performed with the shuffled top-level trajectory.
The identical Metropolis criterion is used --- i.e., (\ref{para_rate}) or its PT analog.
If the exchange attempt is successful, the new lower-level trajectory is continued from the 
accepted configuration, and the top-level trial configuration is simply discarded.
The process is repeated as long as necessary.

Pseudo-exchange processes are useful for several reasons:
(i) ps-ex processes may be used with any exchange simulation;
(ii) much lower acceptance ratios are still efficient because frequent pseudo-exchange 
attempts are inexpensive in a serial scheme; and
(iii) because of the weaker acceptance ratio requirements, larger gaps among ladder 
levels (e.g., $T$ increments in PT) can be tolerated.

 
\begin{figure}[tp]
\begin{center}
\vspace{-.1cm}
\begin{minipage}{\columnwidth}
  \epsfig{file=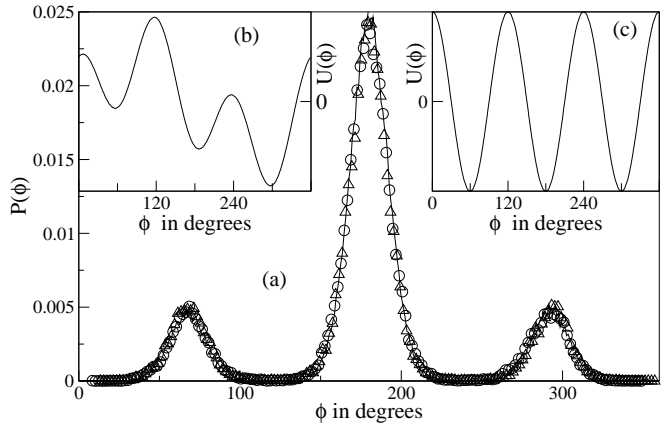, width=\columnwidth}
    \vspace{-.9cm}
\end{minipage}
\end{center}
\caption{The res-ex algorithm produces canonical sampling 
despite a poor coarse-grained potential. 
(a) Probability densities $P(\phi)$ for butane. Reference 
data obtained from standard simulation are indicated by the 
solid line. The res-ex simulation 
with the asymmetric potential in (b) is plotted with triangles, and the res-ex 
simulation with the symmetric potential in (c) is plotted with circles.}
\vspace{-.5cm}
\label{butanefig}
\end{figure}
 
\emph{Simple res-ex demonstration: Butane.} We present simulation results for a single 
butane molecule, in order to show that the res-ex/ps-ex algorithm does indeed 
reproduce the correct 
ensemble. For this purpose, butane is ideal, as it is small enough to permit 
generation of a converged ensemble by standard methods, which is then easily visualized by 
projecting onto a single reaction coordinate.
 
Results are presented for
two different low-resolution potentials, one of which intentionally breaks the symmetry
between the two gauche conformers.
Comparison data were generated with the TINKER v.\ $4.2$ simulation package\cite{tinker}, 
using the CHARMM-27 all-atom force field\cite{charmm}. 
The all-atom butane molecule was simulated in vacuum for 
$1$ $\mu$sec, at a temperature of $298$ K. Langevin dynamics were used, with a friction 
coefficient of $91$ ps$^{-1}$. The 
same force field, dynamics, and simulation package were used for the 
high-resolution portions of the resolution exchange run, except that 
every $10$ fsec a res-ex move was attempted. $10^{5}$ resolution exchanges 
were attempted, for a total  
trajectory length of $1$ nsec. 

FIG.\ref{butanefig} compares the results of two different $1$ nsec res-ex simulations to a 
$1$ $\mu$sec reference trajectory. Plotted is the distribution of the 
C-C-C-C dihedral, $\phi_{1}$, which measures the populations of the three conformers. 
The res-ex simulations reproduce the equilibrium distribution, as measured in the
comparison simulation, regardless of the potential used for the low-resolution
simulation.
The low-resolution model was a one-dimensional potential of the 
form $A\cos (3\phi)+B\sin(\phi)$, where $\phi$ is
the C-C-C-C dihedral. The asymmetric potential, shown in (b), has $A=B=1$, 
while the symmetric potential in (c) has $A=1$ and $B=0$.

\emph{Res-ex for a peptide: Dileucine peptide.} We also tested the res-ex/ps-ex method on the dileucine 
peptide (ACE-(Leu)$_{2}$-NME; ``leucine dipeptide''). Though still a long way 
from a full size protein, $50$ atom dileucine allows us to address a number of 
issues which need to be considered before tackling a full size protein.
A united atom (UA) representation, which omits nonpolar hydrogens, is a 
natural choice for the low-resolution model.
For dileucine, this reduces the number of atoms from $50$ in an all-atom (AA) representation to $24$ in UA.

\begin{figure}[bp]
\epsfig{file=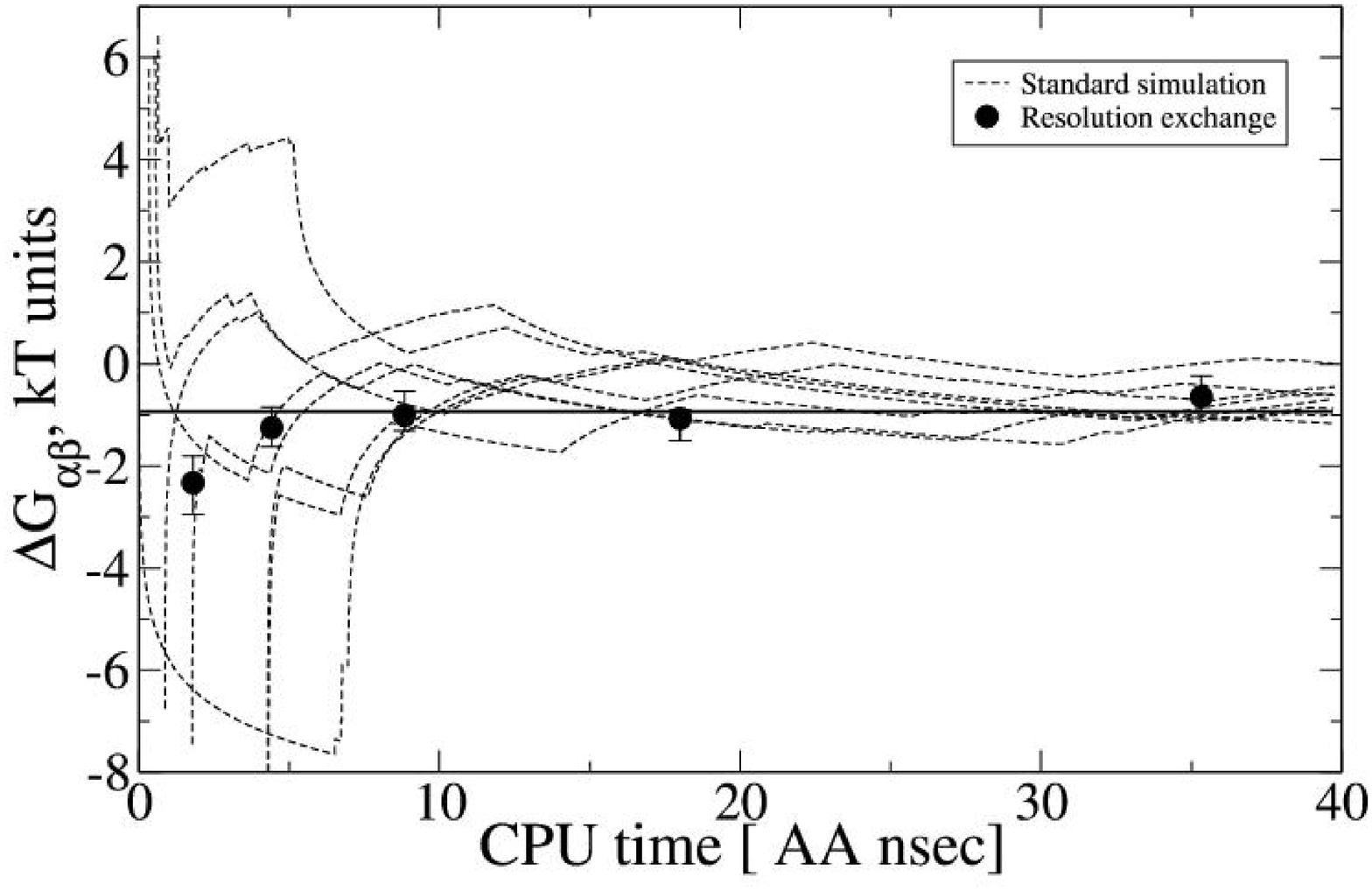, width=\columnwidth}
\caption{Res-ex simulation accelerates equilibration among dileucine peptide substates. 
Dashed lines show running estimates of the inter-substate 
$\Delta\text{G}_{\alpha\beta}$ as a function of time for 
$8$ independent, $40$ nsec trajectories, and the solid line shows the  
estimate of $\Delta\text{G}_{\alpha\beta}$ from $600$ nsec of standard simulation. 
Each symbol with error bars gives the average and range 
of $8$ independent resex simulations, displaced from the origin to reflect total CPU time (measured in 
all-atom timesteps), including 
investment in the united atom simulation. The efficiency gain can be estimated by the relative ranges of 
the $\Delta\text{G}_{\alpha\beta}$ estimates.}
\vspace{-0.3cm}
\label{fig3}
\end{figure}
The goal is to generate efficiently a converged ensemble of conformations for all-atom 
dileucine, using the ps-ex/res-ex protocol. We assess convergence by considering 
the free energy difference between the two dominant conformations, 
distinguished by rotations about 
the $\psi_{1}$ angle of leu$1$ and the $\phi_{2}$ angle of leu$2$. 
Transitions between these two basins are hampered by a significant barrier, and 
therefore occur rarely (approx. $1/3$ nsec$^-1$ at $300$ K for AA). 
We define the ``$\alpha$'' conformations by $-105<\psi_{1}<0$ and $-145<\phi_{2}<-25$, 
and the ``$\beta$'' conformations by $30<\psi_{1}<-155$ and $-160<\phi_{2}<-40$. 

The AA dileucine molecule was modelled with the OPLSaa force field\cite{oplsaa}. For 
UA dileucine, we used a slightly modified version of the OPLSua force 
field\cite{oplsua}, altering a few of the bond length and bond 
angle parameters to match those in the all-atom force field: these simple 
changes reduce the likelihood of exchange-induced steric clashes.
Both simulations were carried out with the TINKER v.\ 4.2 
simulation package\cite{tinker}, 
using Langevin dynamics with a $91$ psec$^{-1}$ friction coefficient, a $1$ fsec integration timestep 
and GB/SA implicit solvation\cite{still-gbsa}. 

Implementing the res-ex/ps-ex strategy, we first carried out a
simulation of united-atom dileucine (FIG.\ref{fig1}(a)). We then randomly reshuffled 
this trajectory (FIG.\ref{fig1}(b)) in order to generate a pseudo-trajectory with much more 
frequent $\alpha\leftrightarrow\beta$ transitions than are 
observed in the original trajectory. This 
randomized trajectory (still Boltzmann distributed) is then used 
to generate the all-atom trajectory (FIG.\ref{fig1}(c)) 
via the res-ex protocol. Notice that $\alpha\leftrightarrow\beta$ transitions are observed far 
more frequently in the all-atom trajectory with exchange (about $30/$nsec) than without 
(about $1/3$ nsec, FIG.\ref{fig1}(d)).  

Assessing convergence and efficiency of a protein simulation are generally  difficult tasks. 
Fortunately, the situation here is relatively simple, as we can consider the free energy difference 
between the $\alpha$ and $\beta$ substates ($\Delta\text{G}_{\alpha\beta}$). 
In FIG.\ref{fig3}, $8$ res-ex simulations are compared to $8$ standard stochastic dynamics 
simulations. Each res-ex data point represents the average and range of $8$ 
independent res-ex trajectories, 
with res-ex moves attempted every $10$ fsec. 
The res-ex estimates 
are displaced from the origin to reflect the time invested in the united atom model. 
We allotted AA simulation time based upon matching precision among levels---i.e., generating 
the same number of inter-basin hops as in the UA simulation.
Convergence is assessed by considering the spread among the independent simulations.

It is clear that the res-ex simulations reproduce the $\Delta\text{G}_{\alpha\beta}$ estimated 
from the standard simulations. This is accomplished despite the failure of the united atom 
model to reflect correctly the populations of the $\alpha$ and $\beta$ states: 
$\Delta\text{G}_{\alpha\beta}$(united atom) = $0.690 k_{B}T$, with the wrong sign. 
This is an important point, as it is 
known that united atom models do not reproduce all atom behavior\cite{freed-uavaa}. 
More important for res-ex simulation is that the coarse-grained 
model explores conformational space 
more rapidly, as well as being ``exchangeable'' with the more detailed model. 

It is further clear that the res-ex results are generated with significantly higher 
efficiency. For a given amount of CPU time 
(nsec in FIG.\ref{fig3}), the res-ex estimates exhibit high accuracy 
with a greatly reduced uncertainty. For example, $5$ nsec of resolution exchange 
simulation generated an estimate for $\Delta\text{G}_{\alpha\beta}=-1.25\pm0.40 k_{B}T$, while
$75$ nsec of standard simulation are required to reach a comprable level of accuracy 
and precision, indicating a $15$-fold savings in CPU time. 
We emphasize that our analysis gives a true efficiency estimate,
since it includes the total 
CPU time, rather than the cost for one of many parallel simulations.  

The acceptance ratio of attempted pseudo
exchange moves need not be $20\%$, as conventional wisdom dictates\cite{kofke-accep}. Indeed,
even a very small fraction of accepted exchanges can greatly enhance efficiency, provided
those exchanges generate novel conformations. The goal is to optimize diffusion in
conformation space, not acceptance ratio. In res-ex 
trajectories presented here the  
average acceptance ratio was only $0.156\%$. Nonetheless, high efficiency is obtained 
because successful exchanges with a shuffled top-level trajectory are very likely 
to generate novel conformations, at a fraction of the cost of standard simulation.  

\emph{Discussion.}
We have introduced two extensions of parallel-tempering/replica-exchange which show 
promise for improved efficiency of biomolecular simulations.
``Resolution exchange'' enhances sampling of an expensive, high-resolution model using 
a cheaper, coarse-grained model with the resolution exchange protocol. 
Generalization to a ladder of models is formally trivial. The sampling in the 
high-resolution model satisfies detailed balance, and therefore generates an equilibrium 
ensemble. 
The further introduction of the ``pseudo-exchange'' process permits 
the bulk of computer resources to be 
invested in sampling and crossing barriers at the top level of the exchange ladder (highest 
temperature or lowest resolution), and only incremental additional simulation is required at lower levels.

The treatment of even larger, more complex molecules will be the subject of future research. 
We emphasize that our efficiency gains were obtained using only a two-level ladder, implying that much 
greater efficiency is possible with additional levels --- which can be added at 
small cost via pseudo-exchange.
A long-term goal is to develop a full ladder of reduced, exchangeable models, 
extending up to the ``united residue'' level, because even UA 
computations require long simulation times.
For temperature-based exchange, a PT implementation of the pseudo-exchange 
approach can be applied to explicitly solvated 
proteins with only a minimum of modification to common software packages.
Our own preliminary PT/ps-ex simulations indicate, interestingly, that successful dileucine exchanges 
can easily be implemented with temperature gaps of $200$ K or more.
Ultimately, resolution and temperature 
exchange might be combined for high-efficiency simulations of biomolecules.

Two limitations should be mentioned. First, we do not expect the present 
algorithm to enable exchange between continuum and explicit solvent representations. 
However, the present degree of undersampling of proteins, when using
continuum solvent representations, warrants pursuit of this problem in 
its own right. A second limitation is that, to be exchangeable, two models 
must be sufficiently ``similiar'': there should be overlap between low-energy 
coarse variable conformations. Yet, a process of incremental 
coarsening---changing part of a molecule at a time, and which we have 
already implemented for dileucine (data not shown)---
will minimize this difficulty for larger systems.  

\emph{Acknowledgements. }We thank Carlos Camacho and Bob Swendsen for 
fruitful discussions and comments on the manuscript. This work was supported 
by the Department of Environmental and Occupational Health and the Department of Computational 
Biology at the University of Pittsburgh, and by the NIH (grants ES007318 and GM070987). 

   

\end{document}